\documentclass[10pt] {article}
\usepackage{times}
\usepackage{graphics}
\usepackage{makeidx}
\usepackage{latexsym}
\usepackage{bm}
\usepackage{amsmath}
\usepackage{amsthm}
\usepackage{amssymb}
\makeindex

\newcommand{\beq}{\begin{equation}}
\newcommand{\eeq}{\end{equation}}

\newcommand{\ilpdif}[2]{\partial #1/{\partial #2 }}
\newcommand{\pdif}[2]{\frac{\partial #1}{\partial #2}}
\newcommand{\pddif}[3]{\frac{\partial^2 #1}{\partial #2 \partial #3}}

\newcommand{\defn}{\begin{quote}{\bf Definition. }}
\newcommand{\edefn}{\end{quote}}
\newcommand{\thm}{\begin{theorem}}
\newcommand{\ethm}{\end{theorem}}

\newcommand{\bmat}[1]{\left ( \begin{array}{#1}}
\newcommand{\emat}{\end{array}\right )}

\newcommand{\ts}{^{\sf T}} 

\newcommand{\X}{{\bf X}}

\newcommand{\A}{{\bf A}}
\newcommand{\B}{{\bf B}}
\newcommand{\Bt}{{\bf B}\ts}

\newcommand{\p}{{\bf P}}
\newcommand{\K}{{\bf K}}

\newcommand{\Tk}{{\bf T}_k}
\newcommand{\Tm}{{\bf T}_m}
\newcommand{\Tkm}{{\bf T}_{km}}
\newcommand{\Gi}{{\bf G}^{-1}}

\newcommand{\Bs}{{\bf P}{\bf K}\ts}
\newcommand{\Bts}{{\bf KP}\ts}
\newcommand{\As}{{\bf KK}\ts}
\newcommand{\Gis}{{\bf PP}\ts}

\newcommand{\bp}{{\bm \beta}}

\newcommand{\tr}[1]{{\rm tr}\left ( {#1} \right )}
\newcommand{\diag}[1]{{\rm diag} \left ( {#1} \right )}
\theoremstyle{definition}

\theoremstyle{plain}
\newtheorem{theorem}{Theorem}

\newcommand{\eps}[3]
{{\begin{center}
 \rotatebox{#1}{\scalebox{#2}{\includegraphics{#3}}}
 \end{center}}
}

\begin{document}
\title{Fast stable direct fitting and smoothness selection \\ for Generalized Additive Models}
\author{ Simon N. Wood\\  Mathematical Sciences, University of Bath, Bath BA2 7AY U.K.\\{\tt s.wood@bath.ac.uk}}

\maketitle

\begin{abstract}
Existing computationally efficient methods for penalized likelihood GAM fitting employ iterative smoothness selection on working linear models (or working mixed models). Such schemes fail to converge for a non-negligible proportion of models, with failure being particularly frequent in the presence of concurvity. If smoothness selection is performed by optimizing `whole model' criteria these problems disappear, but until now attempts to do this have employed finite difference based optimization schemes which are computationally inefficient, and can suffer from false convergence. This paper develops the first computationally efficient method for direct GAM smoothness selection. It is highly stable, but by careful structuring achieves a computational efficiency that leads, in simulations, to lower mean computation times than the schemes based on working-model smoothness selection. The method also offers a reliable way of fitting generalized additive mixed models.

\bigskip


\end{abstract}

\renewcommand{\thefootnote}{\fnsymbol{footnote}}

\section{Introduction}

\begin{figure}
\eps{-90}{.32}{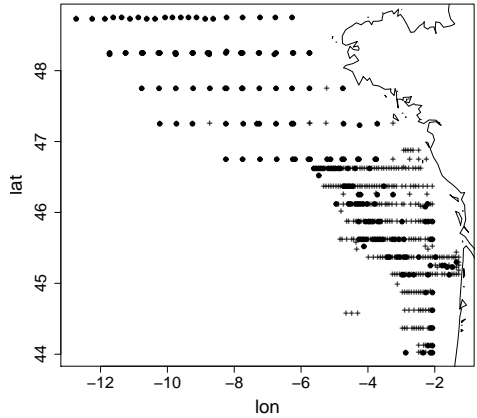}

\vspace*{-.5cm}

\caption{ Presence ($\bullet$) absence ($+$) data for mackerel eggs off the west coast of France. The data are based on the larger data set discussed in section \ref{mack.section}. There are substantial problems fitting a GAM to these data with existing methods.
\label{mack.bin.data}}
\end{figure}

\begin{figure}
\eps{-90}{.4}{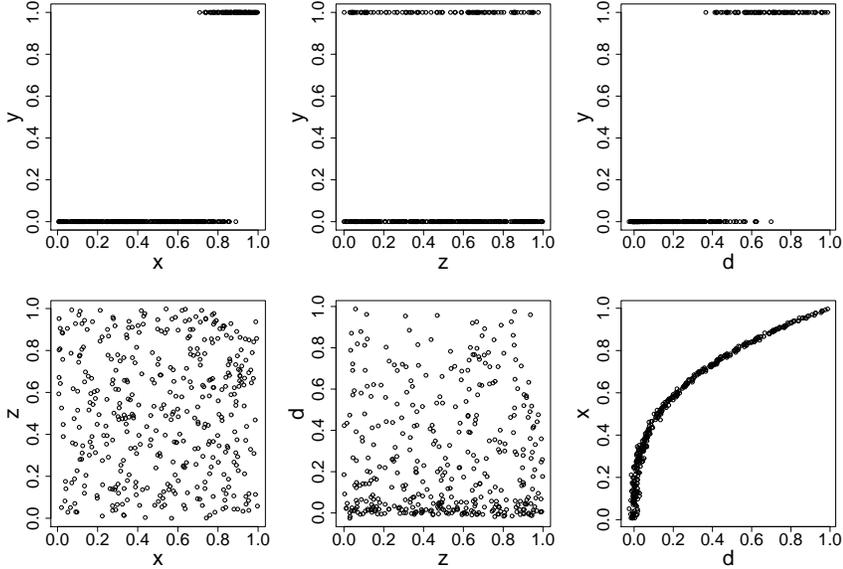}

\vspace*{-.5cm}

\caption{ Simulated data with a serious concurvity problem. Response $y$ depends on covariates $z$, $x$ and $d$. Existing methods have difficulty fitting an appropriate GAM to these data.
\label{concurvity.data}}
\end{figure}

There are three basic approaches to estimating the smoothness of Generalized Additive Models within a penalized likelihood framework. The first is to develop an efficient GCV (Craven and Wahba, 1979) or AIC (Akaike, 1973) based smoothness selection method for the simple (i.e. non-generalized) additive model case (e.g Gu and Wahba, 1991, for smoothing spline ANOVA models; see also Mammen and Park, 2005, for backfit GAMs), and then to apply this method to each working additive model of the penalized iteratively re-weighted (P-IRLS) scheme used to fit the GAM. This was initially proposed by Gu, (1992) for generalized smoothing spline ANOVA models: he termed the approach `Performance Oriented Iteration'. Wood (2000) extended the method to computationally efficient low rank GAMs based on penalized regression splines, while Wood (2004) developed it further by providing an optimally stable smoothness selection method for simple additive models. The second approach is to treat the GAM as a generalized linear mixed model (see e.g. Ruppert et al. 2003), so that smoothing parameters become variance components. In the non-generalized case these variance components can be estimated by maximum likelihood or REML, but in the generalized case methods based on iterative fitting of working linear mixed models are used (e.g. Breslow and Claytons's, 1993, `PQL' approach).  There can be no guarantee of convergence for methods based on iteratively selecting the smoothness of working linear (mixed) models, so the third approach avoids this problem by selecting smoothing parameters directly, based on AIC or GCV  for the actual model: it goes back at least as far as  O'Sullivan et al. (1986). Gu's (2004) {\tt gss} package has an implementation of this, based on work by Kim and Gu (2004), and Wood (2006) attempted to extend an earlier performance oriented iteration method (Wood, 2004) in this direction. The difficulty with the direct approach is that its extra non-linearity makes efficient and stable methods difficult to develop: in consequence, the Gu (2004) and Wood (2006) methods are based on inefficient finite differencing based optimization, which is not always reliable. Figures \ref{mack.bin.data} and \ref{concurvity.data} show data sets for which these three existing approaches fail.

Figure \ref{mack.bin.data} shows data on the presence or absence of Mackerel eggs from a sub region of a survey where absences are sufficiently common that it might be worthwhile to model presence/absence before attempting to model abundance, the real quantity of interest. Covariates longitude, latitude, distance from the continental shelf edge, sea bed depth, surface temperature and temperature at 20m depth are available, and in the absence of more detailed prior knowledge a generalized additive model:
\beq
{\rm logit}\{E(p_i)\} = f_1({\tt lon}_i,{\tt lat}_i) + f_2({\tt c.dist}_i) + f_3({\tt b.depth}_i) + f_4({\tt t.surf}_i) + f_5({\tt t.20m}_i) \label{mack.logistic}
\eeq
might be appropriate, where $p_i$ is 1 for presence and 0 for absence of eggs. A Bernouilli distribution for $p_i$ is assumed. A tensor product of two rank 8 penalized regression splines should be more than sufficiently flexible for $f_1$, while the remaining terms can be represented by rank 10 penalized regression splines. Performance oriented iteration diverges when attempting to fit this model. PQL based model fitting either fails to converge or `converges' to a clearly sub-optimal `completely smooth' model, depending on numerical details (if treated in the same way as a standard penalized likelihood based model, the completely smooth model would have an AIC around 84 larger than the AIC optimal model). Direct optimization of the whole model (generalized) AIC also fails, using either a pure finite difference based optimization, or Wood's (2006) approach based on finite differencing only for second derivatives. In these cases careful examination of the optimization results does indicate the possibility of problems, and estimated minimum AICs are approximately 20 above the actual optimum. All computations were performed using R 2.4.0. with GAM setup based on the {\tt mgcv} package and mixed model estimation based on the {\tt nlme} and {\tt MASS} packages. Direct optimization was performed using the {\tt nlm} routine (results from {\tt optim} generally being substantially more problematic).

Data sets that cause problems for existing GAM estimation and smoothness selection methods are not hard to find, and the root cause is often some form of concurvity (i.e. the presence of covariates which are themselves well modelled as smooth functions of other covariates). Often such problems are difficult to isolate, but figure \ref{concurvity.data} shows simulated data for which the problem is clear. PQL and POI based methods both fail when used to fit the model:
\beq
{\rm logit}\{E(y_i)\} = f_1(x_i,z_i) + f_2(d_i),~~~ y_i \sim {\rm  Bernoulli}. \label{conc.model}
\eeq
to these data. $f_1$ and $f_2$ were represented using thin plate regression splines of basis dimension 30 and 10, with additional shrinkage so that a large enough smoothing parameter can shrink the function to zero (see Wood, 2006, 4.1.6). PQL diverges until the linear mixed model estimator fails, while POI simply fails to converge. Again, direct smoothness selection using general purpose optimizers and finite difference derivatives fails (substantially) to locates the AIC optimal model. This latter failure occurs whether or not extra help in the form of exact first derivatives is supplied.

It should be noted that the impact of concurvity in these examples is different in kind to the well publicized difficulties discussed by Ramsay, Burnett and Krewski (2003) in which concurvity can cause backfitting approaches to GAM estimation (as in Hastie and Tibshirani, 1990) to substantially underestimate estimator variances. For the direct GAM fitting approach, discussed here, the issue is reliably estimating the model in the presence of concurvity driven ill-conditioning. Once the model is estimated the corresponding variance estimates will automatically take into account the effect of the concurvity, so that variance correction approaches of the sort discussed, for example, in Figueiras, Roca-Pardi\~nas and Cadarso-Su\'arez (2005) are not needed (and might actually be counter productive). In other words, if the computational difficulties caused by concurvity can be solved for the direct fitting approach, then we avoid the major part of concurvity driven variance bias `for free'.

For applied use of GAMs these convergence failures are obviously problematic and the aim of this paper is to develop methods that eliminate them to the maximum extent possible. General fitting methods for GAMs should simply work (much as we expect of GLM fitting routines), without the need for tuning, otherwise the fitting methods get in the way of practical modelling. So the objective here is to produce the most reliable method possible for penalized likelihood based GAM estimation with AIC or GCV type smoothness selection. This suggests:
\begin{enumerate}
\item The method must be `direct' in the sense already defined, so that fixed optimum of the smoothness selection criteria exist.
\item For maximal reliability, optimization of the criteria should be based on a full Newton method, utilizing exact first and second derivatives, not approximations.
\item The method must be able to deal with any linear degeneracy in the model, such as that caused by severe concurvity, or by the heavy smoothing that is appropriate in many practical modelling situations.
\end{enumerate}
In addition the method must meet the practical consideration of being computationally efficient, and, given the non-linearities imposed by the direct approach, point 2 presents the main challenge in this regard. None of the forgoing aims are very difficult to achieve if the method is allowed an operation count that is cubic in the number of data, but such an expensive scheme would be of no practical interest.

Faced with the task of producing such a method it might be tempting to abandon penalized likelihood in favour of a fully Bayesian MCMC approach (e.g. Fahrmeir and Lang, 2001; Fahrmeir, Kneib and Lang, 2004 or Brezger and Lang, 2006), but this is not always the answer, and can make problems harder to detect. Firstly, convergence problems can often become mixing problems. For example, using the data in figure \ref{mack.bin.data}, MCMC simulation with the mackerel model (\ref{mack.logistic}) gives markedly reduced acceptance rates (minimum down to 10\% from 75\%), increased between chain variability and appears to require increased burn in, relative to similar models for data simulated without serious concurvity problems. Computations were performed using the state of the art BayesX package (Brezger, Kneib and Lang, 2007), so the model representation was slightly different to that used for the other computations in that the Bayesian P-splines of Lang and Brezger (2004) were used to represent the smooths components. The second issue with MCMC methods is that computational feasibility requires that the prior covariance matrix (smoothing penalty matrix) is sparse (this is the consideration that drives the choice of P-splines in BayesX). Many practically useful smoothers do not have this property (e.g. thin plate splines, as used in model (\ref{conc.model})). In the absence of sparsity then the computational cost is of the order of the cube of the largest (non-sparse) smoothing basis dimension, multiplied by the chain length. For smooths with a simple prior covariance matrix structure, then in principle this cost can be reduced to the {\em square} of the largest basis dimension, multiplied by the number of steps of the chain actually used for posterior inference.  Some form of Demmler-Reinsch orthogonalization  (see e.g. Wood, 2006, 4.10.4) is what is needed to achieve this. However, such orthogonalization has yet to be tried in practice, may lead to more difficulties in setting the hyper-prior on smoothing parameters, and can not be done at all for the kind of penalty required in order to ensure scale invariance in smooth interaction terms (e.g. Wood, 2006, 4.1.8 or Wahba 1990, Chapter 10). Situations in which quasi-likelihood is appropriate are also awkward to handle in an MCMC context. On the other hand the Bayesian MCMC approach improves what can be done with non smooth random effects, and is usually the best option when large numbers of such random effects are required.

The remainder of this paper is structured as follows. Section 2 reviews essential background and discusses smoothness selection criteria. Section 3 proposes a method for efficient and stable optimization of such criteria, and hence for GAM fitting. Section 4 illustrates the comparative performance of the new method using simulated and real data, including a GAMM example.

\section{GAM estimation and selection \label{section.gam.fit}}

Generalized additive models (Hastie and Tibshirani, 1986) are generalized linear models (Nelder and Wedderburn, 1972) in which the linear predictor is partly composed from a sum of smooth functions of some or all of those covariates. Hence the basic model structure is
\beq
g\{E(y_i)\} = \X_i^*{\bm \theta} + \sum_j f_j(x_j) \label{a.gam}
\eeq
where the $y_i$ are observations on independent random variables from some exponential family distribution, or failing that, have a mean variance relationship appropriate for use of a quasi-likelihood approach to inference. $g$ is a smooth monotonic `link' function. $\X_i^*$ is the $i^{\rm th}$ row of the model matrix for any strictly parametric model components, and $\bm \theta$ is the corresponding parameter vector. The $f_j$ are smooth functions of covariates $x_j$, which may be vector covariates. The $f_j$ are subject to identifiability constraints, typically that $\sum_i f_j(x_{ji})=0~\forall~j$. Sometimes each smooth function may also be multiplied by some covariate, yielding a `variable coefficient' model (Hastie and Tibshirani, 1993): the extension is trivial to handle in practice (as recognized by Eilers and Marx, 2002; it has also been available in R package {\tt mgcv} since early 2002). The model can further be extended to include extra random effect terms to arrive at the generalized additive mixed model (GAMM, eg. Lin and Zhang, 1999). The link between penalized regression and mixed modelling that lets GAMs be estimated as GLMMs also means that GAMMs can be estimated by the methods discussed here (see section \ref{gamm.sim}).

The first step in GAM estimation is to represent the smooth terms in (\ref{a.gam}) using bases with associated penalties (see, e.g., Marx and Eilers, 1998; Wood, 2006). Each smooth term is represented as
$$
f_j(x_j) = \sum_{k=1}^{K_j} \beta_{jk} b_{jk}(x_j)
$$
where the $b_{jk}(x_j)$ are known basis functions, chosen to have convenient properties, while the $\beta_{jk}$ are unknown coefficients, to be estimated. Associated with each smooth function is one or more measures of function `wiggliness' ${\bm \beta}_j\ts \tilde {\bf S}_j {\bm \beta}_j$, where $\tilde {\bf S}_j$ is a matrix of known coefficients. Typically the wiggliness measure evaluates something like the univariate spline penalty $\int f_j^{\prime\prime}(x)^2dx$ or its thin-plate spline generalization, but it may also be more complex, such as tensor product smooth penalty with multiple ${\bm \beta}_j\ts \tilde {\bf S}_j {\bm \beta}_j$ terms, (e.g. Wood, 2006, section 4.1.8). Intermediate rank basis- penalty smoothers of this sort go back at least as far as Wahba (1980) and Parker and Rice (1985). Hastie and Tibshirani (1990, section 9.3.6) discussed using them for GAMs and O'Sullivan (1986) demonstrated their use in a wide variety of problems.

Given bases for each smooth term, the GAM, (\ref{a.gam}), can be re-written as a GLM, $g\{E(y_i)\} = \X_i \bp$, where $\X$ includes the columns of $\X^*$ and columns representing the basis functions evaluated at the covariate values, while $\bp$ contains ${\bm \theta}^*$ and all the  smooth coefficient vectors, $\bp_j$. The fit of this GLM is most conveniently measured using the deviance:
$$
D(\bp) = 2 \{l_{\rm max} - l(\bp)\}\phi
$$
where $l$ is the log-likelihood, or log-quasi-likelihood of the model, and $l_{\rm max}$ is the maximum possible value for $l$ given the observed data, which is obtained by considering the MLE of a model with one parameter per datum (under which the model predicted $E(y_i)$ is simply $y_i$). $\phi$ is a scale parameter, and the definition of $D$ means that it can be calculated without knowledge of $\phi$.  Maximizing the (quasi-) likelihood is equivalent to minimizing the deviance, and in several ways the deviance behaves rather like the residual sum of squares in linear modeling (see McCullagh and Nelder, 1989 for further details).

If the bases used for the smooth functions, $f_j$, are large enough to be reasonably sure of avoiding mis-specification, then the model will almost certainly overfit if it  is estimated by minimizing the deviance. For this reason GAMs are estimated by minimizing
$$
D(\bp) + \sum_j \lambda_j {\bp \ts}{\bf S}_j \bp
$$
where the $\lambda_j$ are smoothing parameters and the ${\bf S}_j$ are the $\tilde {\bf S}_j$ suitably padded with zeroes so that $\bp \ts {\bf S}_j\bp = {\bm \beta}_j\ts \tilde {\bf S}_j {\bm \beta}_j$. For later notational convenience, define ${\bf S} = \sum_j \lambda_j {\bf S}_j$. The $\lambda_j$ control the smoothness of the component smooth functions. Smoothness selection is about choosing values for the $\lambda_j$.

Given values for the $\lambda_j$, the penalized deviance can be minimized by penalized iteratively re-weighted least squares (P-IRLS, see e.g. Wood, 2006, for one derivation, and Green and Silverman, 1994, for more information on penalized likelihood and GLMs). Let $V(\mu)$ be the function such that ${\rm var}(y_i) = V(\mu_i) \phi $. $V$ can be written down for all exponential family distributions, and is always available if using quasi-likelihood. Let $\omega_i$ denote any prior weights on particular data points (used to weight the component of deviance attributable to each datum). Then iterate the following steps to convergence.
\begin{enumerate}
\item Using the current $\mu_i$ estimate, evaluate the weights, $w_i = \omega_i^{1/2}V(\mu_i)^{-1/2}/g^{\prime}(\mu_i)$, and the pseudodata,
$z_i = g^\prime(\mu_i)(y_i - \mu_i) + \eta_i$, where $\eta_i = g(\mu_i)$ (the `linear predictor').
\item Let $\bf W$ be the diagonal matrix of $w_i$ values. Minimize the penalized least squares objective
\beq
\| {\bf W}({\bf z} - \X\bp) \|^2 + \sum_j \lambda_j {\bp \ts}{\bf S}_j \bp \label{work.obj}
\eeq
w.r.t. $\bp$ to find the next estimate of $\bp$, and hence of ${\bm \eta} = \X\bp$ and $\mu_i = g^{-1}(\eta_i)$.
\end{enumerate}
The iteration can be initialized by setting $\hat \mu_i = y_i$ (with adjustment to avoid infinite $\hat \eta_i$). Divergence is rare, but can be dealt with by step halving (provided an MLE exists). At convergence the parameter estimates, $\hat \bp$, minimize the penalized deviance.

Note that this direct fitting approach makes it straightforward to directly estimate coefficient variances (see e.g. Wood, 2006, section 4.8) thereby completely sidestepping the well publicized problem of concurvity driven variance underestimation, that can affect backfitting methods of GAM fitting (see Ramsay, Burnett and Krewski, 2003, for example).

\subsection{Smoothness selection}

Performance oriented iteration (POI) uses GCV or Mallows' $C_p$ (Mallows, 1973) applied to each fitting problem (\ref{work.obj}) in order to select smoothing parameters. This often converges to fixed $\hat \bp, \hat {\bm \lambda}$, but less often it diverges, or cycles, with failure being particularly frequent for binary data (see section \ref{examples.section} or the Introduction). Mixed model alternatives such as PQL are no better. An alternative, which avoids this fundamental convergence issue, is to base smoothness selection on a criterion applied to the GAM itself and evaluated at convergence of the P-IRLS.

If $\tau$ denotes the effective degrees of freedom of the penalized fit, then one could seek to minimize the generalized AIC:
$$
{\cal V}_a({\bm \lambda}) = D(\hat \bp) + 2 \gamma \tau
$$
in the case where $\phi$ is known, or the generalized GCV score
$$
{\cal V}_g({\bm \lambda}) = nD(\hat \bp)/(n - \gamma \tau)^2
$$
otherwise (see Hastie and Tibshirani, 1990, Section 6.9 or Wood, 2006, section 4.5). $\tau= \tr{\A}$ where ${\bf A} = {\bf WX}(\X\ts{\bf W}^2\X + {\bf S})^{-1} \X \ts{\bf W}$ is the `influence matrix' of the fitted model (${\bf W}$ is evaluated at convergence). $\gamma$ is an ad hoc tuning parameter, sometimes increased from its usual value of 1 in order to obtain smoother models than would otherwise be selected ($\gamma$ can itself be chosen automatically by, e.g., 10-fold cross validation, but this will not be pursued further here).

Another alternative, proposed by Xiang and Wahba (1996) and Gu and Xiang (2001) is Generalized Approximate Cross Validation, GACV (see also Gu, 2002, section 5.2.2, for a clear exposition). It was initially derived for the situation in which only the canonical link function is used, but the restriction can be relaxed (in the process providing one possible justification for ${\cal V}_a$ and ${\cal V}_g$). Some modification of Gu and Xiang's (2001) approach is the key, as outlined next.

The basic idea is that the Kullback-Leibler distance depends on the model only through the `predictive deviance' of the model, which can be estimated by some version of leave-one-out cross validation. The resulting leave-one-out cross validation criterion is then replaced with a {\em generalized} cross validation criterion. To this end, first write the model deviance as $D(\hat {\bm \eta}) = \sum D_i(\hat \eta_i)$, where $D_i$ is the contribution to the deviance associated with the $i^{\rm th}$ datum. Now the mean predictive deviance of the model can be estimated by
$$
D_{\rm cv} = \sum_{i=1}^n D_i(\hat \eta^{[-i]}_i)
$$
where $\hat {\bm \eta}^{[-i]}$ is the linear predictor obtained by estimating the model from all the data except the $i^{\rm th}$ datum. Minimization of $D_{\rm cv}$ is an attractive way of choosing smoothing parameters as it seeks to minimize the KL distance between the estimated model and the truth; however it is an impractically expensive quantity to attempt to minimize directly.

To progress, follow Gu and Xiang (2001) and let $\hat {\bm \eta}^{[-i]}$ be the linear predictor which results if $z_i$ is omitted from the working linear fit at the final stage of the P-IRLS. This can be shown to imply that
$
\hat \eta_i - \hat \eta^{[-i]}_i =  (z_i - \hat \eta_i) {A_{ii}}/({1-A_{ii}}).
$
But $z_i - \hat \eta_i = g^\prime(\hat \mu_i) (y_i - \hat \mu_i)$, so
$
\hat \eta_i - \hat \eta^{[-i]}_i = g^\prime(\hat \mu_i) (y_i - \hat \mu_i) {A_{ii}}/({1-A_{ii}}).
$
Now take a first order Taylor expansion
$$
D_i(\hat \eta^{[-i]}_i) \simeq D_i(\hat \eta_i) + \pdif{D_i}{\hat \eta_i} (\hat \eta^{[-i]}_i - \hat \eta_i) = D_i(\hat \eta_i) - \pdif{D_i}{\hat \eta_i} \frac{A_{ii}}{1-A_{ii}}g^\prime(\hat \mu_i) (y_i - \hat \mu_i).
$$
Noting that
$$
\pdif{D_i}{\hat \eta_i} = - 2 \omega_i \frac{y_i - \hat \mu_i}{V(\hat \mu_i) g^{\prime}(\hat \mu_i)},
{\rm ~~ we ~have~~}
D_i(\hat \eta^{[-i]}_i) \simeq D_i(\hat \eta_i) + 2 \frac{A_{ii}}{1-A_{ii}} \omega_i \frac{(y_i - \hat \mu_i)^2}{V(\hat \mu_i)}.
$$
Using the same approximation employed in the derivation of GCV, the individual $A_{ii}$ terms are replaced by their average, $\tr{{\bf A}}/n$, to yield,
$$
D_i(\hat \eta^{[-i]}_i) \simeq D_i(\hat \eta_i) + 2 \frac{\tr{{\bf A}}}{n-\tr{{\bf A}}} \omega_i \frac{(y_i - \hat \mu_i)^2}{V(\hat \mu_i)},
$$
and averaging over the data gives a GACV score
$$
{\cal V}_g^* = D(\hat {\bm \eta})/n + \frac{2}{n} \frac{\tr{{\bf A}}}{n-\tr{{\bf A}}}
\sum_{i=1}^n \omega_i \frac{(y_i - \hat \mu_i)^2}{V(\hat \mu_i)}
= D(\hat {\bm \eta})/n + \frac{2}{n} \frac{\tau}{n-\tau} P(\hat {\bm \eta}).
$$
where $P = \sum_i \omega_i(y_i - \hat \mu_i)^2/V(\hat \mu_i)$ is a `Pearson statistic'.
(The final term on the RHS might also be multiplied by $\gamma \ge 1 $ of course.)
Although the basic motivation and approach comes directly from the cited references, the need to accommodate non-canonical links means that the final score differs a little from Gu and Xiang (2001) in the terms in the final summation.

Notice how ${\cal V}^*_g$ is just a linear transformation of (generalized) AIC, with $\hat \phi = P(\hat {\bm \eta}) /\{n-\tr{{\bf A}}\}$ in place of the MLE of $\phi$, and $\tr{\bf A} $ as the model degrees of freedom: so if $\phi$ is known we might as well use ${\cal V}_a$. Viewed from this perspective there is also no reason not to use $D(\hat {\bm \eta}) /\{n-\tr{{\bf A}}\}$ for $\hat \phi$, in which case, for $\tr{{\bf A}} \ll n $, the resulting criterion would be approximately ${\cal V}_g$. Of course these connections are unsurprising: see Stone (1977).

The next section discusses how best to optimize these criteria with respect to the smoothing parameters.

\section{Stable fitting and $\cal V$ optimization \label{fit.details}}

Optimization of the ${\cal V}$ type criteria is basically hierarchical. The criteria are optimized with respect to the smoothing parameters, with any set of smoothing parameters implying a particular set of coefficient estimates $\hat \bp$, which are found by an `inner' P-IRLS iteration.

The dependence of ${\cal V}_g$, ${\cal V}_g^*$ and ${\cal V}_a$ on the smoothing parameters is via $D(\hat \bp) $, $\tau$ and possibly $P(\hat \bp)$, so that the key to successful $\cal V$ optimization is to obtain first and second derivatives of $D(\hat \bp) $, $\tau$ and $P(\hat \bp)$ with respect to the log smoothing parameters, $\rho_j = \log(\lambda_j)$, in an efficient and stable way (logs are used to simplify optimization, since the $\lambda_j$ must be positive). Given these derivatives, the derivatives of the criteria themselves are easily obtained, and they can be minimized by modified Newton, or quasi-Newton methods to select smoothing parameters (e.g. Gill et al., 1981, Dennis and Schnabel, 1983).

The required derivatives in turn depend on the derivatives of $\hat \bp$ with respect to $\bm \rho$. Conceptually, these can be obtained by differentiating the P-IRLS scheme, and updating derivatives alongside the parameters as the P-IRLS progresses. However, while this is the way to obtain expressions for the derivatives, it is a poor way to arrange the computations (a prototype {\em first} derivative based scheme of this sort is proposed in Wood, 2006, but is built on the method of Wood, 2004, making it both inefficient and difficult to extend to a second derivative scheme). Instead, for fixed $\bm \rho$,
\begin{enumerate}
\item Iterate the P-IRLS scheme to convergence of the $\hat \bp$, ignoring derivatives and using the fastest available stable method for solving each P-IRLS problem.
\item At convergence, with the weights, pseudodata and hence all matrix decompositions now fixed, iterate the expressions for the derivatives of the coefficient, $\hat \bp$, with respect to the log smoothing parameters ${\bm \rho}$ to convergence.
\item Evaluate the derivatives of $\tau = \tr{{\bf A}}$ with respect to $\bm \rho$.
\end{enumerate}

Relative to accumulating derivatives alongside the P-IRLS, the method has a number of advantages. Firstly, the basic matrix decompositions and some other component matrix expressions stay fixed throughout the derivative iteration, reducing the cost of using the most stable decompositions for the derivative calculations, and avoiding re-calculation at each step of the P-IRLS. In addition, fewer steps are typically required for the derivative iteration than for the P-IRLS itself, thereby saving the cost of several derivative system updates, relative to parallel accumulation of derivatives and estimates. Purely practically, the derivative update becomes a separate `add-on' to the P-IRLS iteration, which simplifies implementation. The following subsections explain how the method works in more detail.

\subsection{Iterating for derivatives of $\hat \bp$ \label{section.pirls}}

At any step of the P-IRLS, let ${\bf B} = (\X\ts{\bf W}^2\X + {\bf S})^{-1} \X \ts{\bf W}$ and ${\bf z}^\prime = {\bf Wz}$ so that $\hat \bp = {\bf Bz}^\prime$. By differentiating the P-IRLS presented in section \ref{section.gam.fit}, the following update algorithm results.

\noindent {\bf Initialization:}
${\bf z}$ is fixed at its converged value from the P-IRLS, and its derivatives w.r.t. $\bm \rho$ are initially set to zero. The corresponding initial derivatives of $\hat {\bm \beta}$ are then given by
$$
\pdif{\hat \bp}{\rho_k} = \pdif{{\bf B}}{\rho_k} {\bf z}^\prime
{\rm~~~~and~~~~}
\pddif{\hat \bp}{\rho_k}{\rho_m} = \pddif{\bf B}{\rho_k}{\rho_m} {\bf z}^\prime
$$
where the derivatives of $\bf B$ are evaluated with all the $\Tk$, $\Tm$ and $\Tkm$ terms defined in section \ref{B.deriv} set to zero.

\noindent {\bf Iteration:}
The following steps are iterated to convergence (for all $k,m$, such that $k \ge m$).
\begin{enumerate}

\item Update
$$
\pdif{z_i^\prime}{\rho_k} {\rm ~~and~~} \pddif{z_i^\prime}{\rho_k}{\rho_m}
$$
using the current derivatives of $\hat \bp$, as described in appendix A.
\item Using the ${\bf z}^\prime$ derivative from step 1, the derivatives of $\hat \bp $ are updated as follows,
$$
\pdif{\hat \bp}{\rho_k} = \pdif{{\bf B}}{\rho_k} {\bf z}^\prime + {\bf B} \pdif{{\bf z}^\prime}{\rho_k}
{\rm ~~and~~}
\pddif{\hat \bp}{\rho_k}{\rho_m} = \pddif{\bf B}{\rho_k}{\rho_m} {\bf z}^\prime + \pdif{\bf B}{\rho_k}\pdif{{\bf z}^\prime}{\rho_m} + \pdif{\bf B}{\rho_m}\pdif{{\bf z}^\prime}{\rho_k} + {\bf B} \pddif{{\bf z}^\prime}{\rho_k}{\rho_m}.
$$
Note that while $\bf B$ is fixed, its derivatives will change as the iteration progresses.
\end{enumerate}

Convergence of the iteration would usually be judged by examining convergence of the the first and second derivatives of the deviance with respect to ${\bm \rho}$. Calculation of these is routine, given the derivatives of $\hat \bp$: the expressions are provided in Appendix B, and are also needed for obtaining the derivatives of the smoothness selection criteria themselves.

\subsection{Computing with the derivatives of $\bf B$ \label{B.deriv}}

The $\hat \bp$ derivative update scheme involves derivatives of $\bf B$, which need to be spelled out.
Initially expressions for the derivatives will be obtained in terms of $\bf B$, ${\bf A} = {\bf WXB}$, ${\bf G} = \X\ts{\bf W}^2\X + {\bf S}$ and the diagonal matrices:
$$
{\bf T}_k = {\rm diag}\left (\pdif{w_i}{\rho_k} \frac{1}{w_i} \right ) {\rm ~~~~and~~~~}
{\bf T}_{km} = {\rm diag} \left (
\pddif{w_i}{\rho_k}{\rho_m} \frac{1}{w_i} - \pdif{w_i}{\rho_k} \pdif{w_i}{\rho_m} \frac{1}{w_i^2}
\right )
$$
(see appendix A for the derivatives of $w_i$). Noting that
$
\ilpdif{{\bf G}^{-1}}{\rho_k} = - 2 \B\Tk\Bt - e^{\rho_k}\Gi{\bf S}_k \Gi,
$
the first derivative of $\B$ is
$$
\pdif{\B}{\rho_k} = -2 \B\Tk\A - e^{\rho_k}\Gi{\bf S}_k\B + \B\Tk,
$$
while
\begin{multline*}
\pddif{\B}{\rho_k}{\rho_m} = -2 \pdif{\B}{\rho_m} \Tk \A - 2 \B\Tkm\A - 2 \B\Tk \pdif{\A}{\rho_m} + \pdif{\B}{\rho_m}\Tk + \B \Tkm \\- e^{\rho_k} \left (\pdif{\Gi}{\rho_m} {\bf S}_k \B + \Gi {\bf S}_k \pdif{\B}{\rho_m} \right ) - \delta_m^k e^{\rho_k} \Gi{\bf S}_k\B
\end{multline*}
(where $\delta_m^k=1$ if $m=k$ and 0 otherwise). Direct naive use of these expressions would be both computationally expensive (just forming $\bf A$ has $O(n^2q)$ cost) and potentially unstable, since it would not address the ill-conditioning problems that can be a feature of GAMs, especially when there is concurvity present. It is therefore necessary to develop a way of computing with these derivatives which is both maximally stable and keeps computational costs down to a small multiple of $O(nq^2)$, the leading order cost of the P-IRLS.

There are a number of more or less equivalent starting points for such a method, of which the following two are the most straightforward.

\begin{enumerate}
\item Find the Choleski factor, $\bf L$, such that
$$
{\bf L}\ts{\bf L} = \X {\bf W}^2\X + {\bf S}.
$$
This should actually be performed with pivoting (available in LINPACK, Dongarra et al. 1978), in case of co-linearity problems, and the pivoting applied explicitly to the columns of $\X$. The rank, $r$, of the pivoted Choleski factor $\bf L$ can then be estimated, by finding the largest upper left sub-matrix of $\bf L$ with acceptably low condition number (e.g. Golub and van Loan, 1996, section 5.5.7). $\bf L$ is upper triangular, and efficient and reliable estimation of the condition number of triangular matrices is fairly straightforward following Cline et al. (1979). If rank deficiency is detected then all but the $r$ upper left rows and columns of $\bf L$ are dropped along with the corresponding columns of $\bf X$ (for the current iteration only, of course).
Now define $q \times r$ matrix $\bf P$, the first $r$ rows of which are given by $ {\bf L}^{-1} $ with the remaining rows being zero. Also define $n \times r$ matrix ${\bf K} = {\bf WXL}^{-1}$.

\item A somewhat more stable approach first finds a square root $\bf E$ of $\bf S$, so that ${\bf E}\ts{\bf E} = {\bf S} $. A pivoted Choleski decomposition or (symmetric) eigen-decomposition can be used to do this. Next form the QR decomposition
$$
\bmat{c} {\bf WX} \\ {\bf E} \emat = {\bf QR},
$$
where ${\bf Q}$ is a matrix with orthogonal (strictly orthonormal) columns and $\bf R$ is upper triangular. Again this should be performed with pivoting (Golub and van Loan, 1996), which must subsequently  be applied to the columns of $\X$. An LAPACK routine was used for the decomposition (Anderson et al. 1999). Rank deficiency can be dealt with in exactly the same way as it was for ${\bf L}$. Again the redundant columns of $\bf Q$ and rows and columns of $\bf R$ are dropped. Now let $n \times r$ matrix $\bf K$ be the first $n$ rows of $\bf Q$ so that ${\bf WX} = {\bf KR}$, and define $q \times r$ matrix ${\bf P}$ as the matrix with first $r$ rows given by  ${\bf R}^{-1}$ and remainder packed with zeroes.
\end{enumerate}
$\K$ is formed explicitly, while  ${\bf P}$ can either be formed explicitly or computed with using its definition. $\K$ and $\bf P$ are all that are required from the decompositions for subsequent computation. Although the values taken by these two matrices depend on method of calculation, they are used in the same way irrespective of origin. Of the two calculation methods, the second, QR based, method is usually to be preferred over the first, since it avoids the exacerbation of any numerical ill-conditioning that accompanies explicit formation of $\X {\bf W}^2\X$, and instead is based on a stable orthogonal decomposition. The Choleski based method is faster, by about a factor of 2, but irreducible costs of the second derivative calculations, which are the same for both methods, substantially dilute this advantage in practice. A singular value decomposition (see Golub and van Loan, 1996 or Watkins, 1991) based method is also possible, but is more costly and is not pursued here.

Note that for the most part the pivoting used in either method does not effect the subsequent algorithm: it is simply that quantities such as the estimated coefficient vector and its covariance matrix must have the pivoting reversed at the end of the method. The only exception is the one already mentioned, that the pivoting will have to be applied to the columns of $\X$ before it can be used as part of the derivative updating iteration.

It is now straightforward to show that $\Gi = \p\p\ts$ (strictly a sort of pseudo-inverse if the problem is rank deficient), $\B = \p\K\ts$ and $\A = \K\K\ts$, and some work establishes that

$$
\pdif{\B}{\rho_k} = -2 \Bs\Tk\As - e^{\rho_k}\Gis{\bf S}_k\Bs + \Bs\Tk
$$
and
\begin{multline*}
\pddif{\B}{\rho_k}{\rho_m} =
4 \Bs\Tm\As\Tk\As + 2 e^{\rho_m}\Gis{\bf S}_m \Bs\Tk\As- 4 \Bs\Tm\Tk\As\\
 - 2\Bs\Tkm\As  + 4 \Bs\Tk\As\Tm\As
+ 2 e^{\rho_m}\Bs\Tk\Bts {\bf S}_m\Bs \\
 - 2\Bs\Tk\As\Tm - 2 \Bs\Tm\As\Tk
- e^{\rho_m}\Gis{\bf S}_m \Bs \Tk
 + \Bs\Tm\Tk\\
 + \Bs\Tkm + 2e^{\rho_k} \Bs\Tm\Bts{\bf S}_k\Bs
 + e^{\rho_k}e^{\rho_m}\Gis{\bf S}_m\Gis{\bf S}_k \Bs \\
 + 2 e^{\rho_k} \Gis {\bf S}_k \Bs\Tm\As
 + e^{\rho_k} e^{\rho_m}\Gis{\bf S}_k \Gis {\bf S}_m \Bs
 - e^{\rho_k} \Gis {\bf S}_k \Bs \Tm\\ - \delta_k^m e^{\rho_k} \Gis{\bf S}_k \Bs.
\end{multline*}
By inspection of the preceding two equations, it is clear that, given the one off $O(nq^2)$ start up cost of forming $\K$, the multiplication of a vector by either derivative of $\B$ is $O(nq)$. i.e. the leading order cost of computing the smoothing parameter derivatives of $\hat \bp$ and hence of the deviance or Pearson statistic has been kept at $O(nq^2)$.

\subsection{The derivatives of $\tr{\bf A}$\label{trA.deriv}}

Once the derivative iterations detailed in the previous sections are complete, it is necessary to obtain the derivatives of the effective degrees of freedom $\tr{\bf A}$. These are
$$
\pdif{\tr{\A}}{\rho_k} = \tr{ \Tk \A - 2\A\Tk\A - e^{\rho_k}\Bt {\bf S}_k \B + \A\Tk }
$$
and
\begin{multline}
\pddif{\tr{\A}}{\rho_k}{\rho_m}=2 \tr{\Tkm\A + 2\Tk\Tm\A} - 4 \tr{\Tk\A\Tm\A+\Tm\A\Tk\A} \\-
2 \tr{2\A\Tm\Tk\A+\A\Tkm\A}
 + 8 \tr{\A\Tk\A\Tm\A} + 4 \tr{e^{\rho_m}\A\Tk\B\ts{\bf S}_m\B + e^{\rho_k}\A\Tm\B\ts{\bf S}_k\B}\\
- \tr{2e^{\rho_m}\Tk\B\ts{\bf S}_m\B+2 e^{\rho_k}\Tm\B\ts{\bf S}_k\B + \delta_k^m e^{\rho_k}\B\ts{\bf S}_k\B}
+ 2 e^{\rho_m}e^{\rho_k} \tr{\B\ts {\bf S}_m \Gi {\bf S}_k\B}. \label{trA.2deriv}
\end{multline}

Obviously it would be hideously inefficient to evaluate the terms in (\ref{trA.2deriv}) by explicitly evaluating its various component matrices and then evaluating the traces. Rather, efficient calculation of the trace terms rests on: (i) the fact that evaluation of ${\rm diag}({\bf CH)}$, where ${\bf C}$ is $n \times p$, and ${\bf H}$ is $p \times n$, takes $np$ floating point operations, (ii) $\tr{\bf CH} = \tr{\bf HC}$, (iii) careful choice of what to store and in what order to calculate it and (iv) the use of `minimum column' square roots of the penalty matrices ${\bf S}_m$.  The actual evaluation uses the matrices $\bf K$ and $\bf P$, defined in section \ref{B.deriv}, and is detailed in appendix C. The leading order cost of the evaluation is $Mnq^2/2$ operations, where $M$ is the number of smoothing parameters. This is a considerable saving over finite differencing for second derivatives.

Demonstrating that computing with $\K$ and $\p$  is as computationally efficient as is possible actually requires that the derivatives of $\B$ and $\tr{\A}$ be written out in terms of the original matrix decomposition components, and the most efficient computation of each term then be considered. The minimum set of quantities required for the whole calculation is then assembled, at which point it becomes clear that maximum efficiency can be obtained by computing with $\K$, $\p$. This process is exceedingly tedious, and is omitted here.

\subsection{Optimizing AIC, GCV or GACV criteria}

Given the derivatives of $\tau$,  $D$ and $P$ the derivatives of the $ {\cal V}_g$, ${\cal V}_g^*$ or ${\cal V}_a $ are easily obtained, and the criteria can be optimized by Newton type methods. ${\cal V}_g$, ${\cal V}^*_g$ and ${\cal V}_a$ are indefinite over some parts of the smoothing parameter space, since they flatten out completely at very high or very low $\rho_k$ values. In many modeling situations such regions are unavoidable, since the optimal $\rho_k$ for a term that is not needed in a model {\em should} tend to $\infty$. When taking a full Newton approach such indefiniteness is readily identifiable and addressable using an eigen-decomposition, ${\bm \Xi \bm \Lambda \bm \Xi}\ts $, of the Hessian matrix of ${\cal V}$. Following Gill, Murray and Wright (1981) the Hessian is replaced in Newton's method by ${\bm \Xi} \bar {\bm \Lambda}{\bm \Xi}\ts$ where $\bar{\bm \Lambda}_{ii} = |{\bm \Lambda}_{ii}| $. Since the replacement is positive definite, the resulting modified Newton direction is guaranteed to be a descent direction.
Note that the eigen-decomposition is a trivial part of the total computational burden here. Another way of increasing convergence rates is to only optimize smoothing parameters for which the corresponding  gradient of $\cal V$ is large enough to be treated as unconverged. When the converged parameters are optimized at `working infinity', dropping them from optimization tends to improve the quadratic model underlying the Newton update of the remaining parameters. (Parameters re-enter the optimization if their corresponding gradient becomes large again.)

Alternatively, one can work only with first derivatives, and use the quasi-Newton or Newton type algorithms built into R routines {\tt optim} and {\tt nlm}, for example (see R Core Development Team, 2006 and Dennis and Schnabel, 1983). The associated loss of computational speed is smaller than might be expected, as first derivatives are very cheap to obtain, and indefiniteness produces some degradation in  the convergence rates of the `pure' Newton method. However, as the initial example in the introduction emphasizes, a finite differencing based method will not always be reliable when faced with complex models and strong concurvity effects.

\section{Examples \label{examples.section}}

\subsection{Performance in `straightforward' situations \label{easy.sim}}

A  small simulation study was undertaken to illustrate the method's performance in  non-problematic situations, which should not generate numerical problems and where the data do not display concurvity. The example is adapted from one presented in Wahba (1990, section 11.3).

For each replicate, 400 values for each of 4 covariates, $x_1, \ldots, x_4$, were simulated independently from a uniform distribution on $(0,1)$. The covariates were used to produce a scaled linear predictor of the form $\tilde \eta_i = f_1(x_{1i})+f_2(x_{2i})+f_3(x_{3i})$, where, $f_1(x)=2 \sin(\pi x)$, $f_2(x)=\exp(2x)$ and $f_3(x)=x^{11}\{10(1-x)\}^{6}/5 + 10^{4}x^3(1-x)^{10}$.
Response data, $y_i$, were then generated under one of 4 models. (i) Independent Bernoulli random deviates were generated, taking the value 1 with probability $e^{\eta_i}/(1+e^{\eta_i})$, where $\eta_i = (\tilde \eta_i-5)/2.5$; (ii) independent Poisson random deviates were generated with mean $\exp(\eta_i)$, where $\eta_i=\tilde \eta_i/7$; (iii) independent gamma random deviates were generated with mean $\exp(\eta_i)$, with $\eta_i=\tilde \eta_i/7$ (and scale parameter 1); (iv) independent Gaussian random deviates were generated from $N(\eta_i,4 \eta_i)$ truncated (below) at zero, where $\eta_i = \exp(\tilde \eta_i/6)$.

To each replicate a 4 term generalized additive model
$$
g\{E(y_i)\} = f_1(x_{1i})+f_2(x_{2i})+f_3(x_{3i}) + f_4(x_{4i})
$$
was fitted. The link function, $g$, was the logit for the binary data, and log for the other cases. The $f_j$ were represented using rank 10 thin plate regression splines (Wood, 2003). The correct distribution was assumed for response models (i) to (iii), and for (iv) a quasi-likelihood approach was taken, with the variance assumed proportional to the mean. Each model was estimated by 5 alternative methods. (a) By the method presented in this paper using full first and second derivative information; (b) using the first derivative scheme presented here to optimize GCV/AIC scores using the `nlm' routine from R (this seems to be the fastest and most reliabale R general purpose optimizer for this problem); (c) optimizing the GCV/AIC scores using finite difference based gradients with R optimizer `nlm'; (d) using Gu's (1992) performance oriented iteration as implemented in Wood (2004); (e) representing the model as a mixed model and estimating via Breslow and Clayton's (1993) PQL (using the `nlme' library as the underlying mixed model fitter; Pinheiro and Bates, 2000).
For methods a-d AIC was used for the binary and Poisson cases, and GCV for the other two. GACV was also tried but had marginally worse MSE/ predictive deviance than GCV, and is therefore not reported here.

To measure model fit, 10000 new data were generated from the model concerned, and the fitted model was used to predict the (expected value of the) response variable. The prediction error was measured using the mean deviance of the prediction of the 10000 simulated response data {\em minus} the mean predictive deviance using the known truth. This {\em predictive deviance loss} is therefor zero if the model exactly reproduces the truth. In the quasi case the error model is incorrect, so the predictive deviance is not such a natural measure of performance and the mean square error in predicting the (covariate conditional) means over 10000 independent replicate data was used as the prediction error measure (although in fact the conclusions are no different if predictive deviance is used).

\begin{figure}
\eps{-90}{.4}{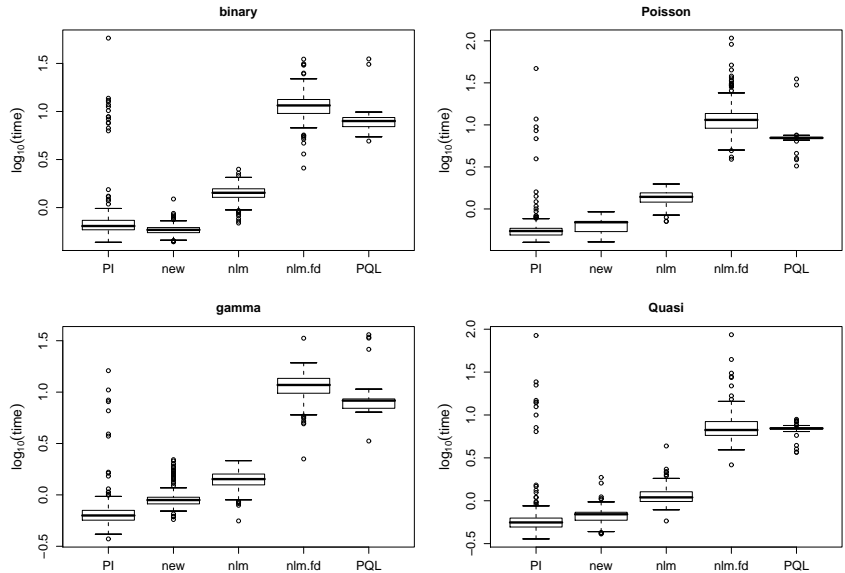}

\vspace*{-.5cm}

\caption{Boxplots of the distribution of the $\log_{10}$(CPU seconds) used to fit the models to the simulated data in section \ref{easy.sim}. `new' is the new method; `nlm' is the new method, but using only first derivatives; `nlm.fd' is the same optimization performed without derivative information; `PI'
is performance oriented iteration; `PQL' is for the GAM estimated as a mixed model. The skew in the PI timing distributions means that the new method has the lowest mean time for all 4 models. The new method is also the most reliable --- see text. \label{timing.fig}}
\end{figure}

 The PQL iteration failed in 20, 8, 13 and 23 out of 200 replicates for the binary, Poisson, gamma and quasi models, respectively, while the POI failed to converge in 13, 5, 5 and 10 out of 200 replicates. The smaller failure rate for performance oriented iteration as opposed to PQL may reflect the fact that the penalized least squares estimator used for POI was specifically designed for use in this application (Wood, 2004). The new method converged successfully for all replicates. The {\tt nlm} based methods produced warnings of potential convergence problems in about 2\% of cases, but none of these in fact appear to be real failures: rather the warnings seem to be triggered by the indefinite nature of the smoothness objective. Failures are of course excluded from the reported predictive deviance (MSE) comparisons and timings.

Time, in CPU seconds, to fit each replicate model was also recorded (on a Pentium M 2.13Ghz processor with a Linux operating system). The results are summarized in figure \ref{timing.fig}. The new method clearly has lower computational cost than all the other methods apart from performance oriented iteration, although it turns out that, for each error model, the {\em mean} time required for the new method is actually less than that required by performance oriented iteration, as a result of the skew in the latter's timing distributions.

\begin{figure}
\eps{-90}{.4}{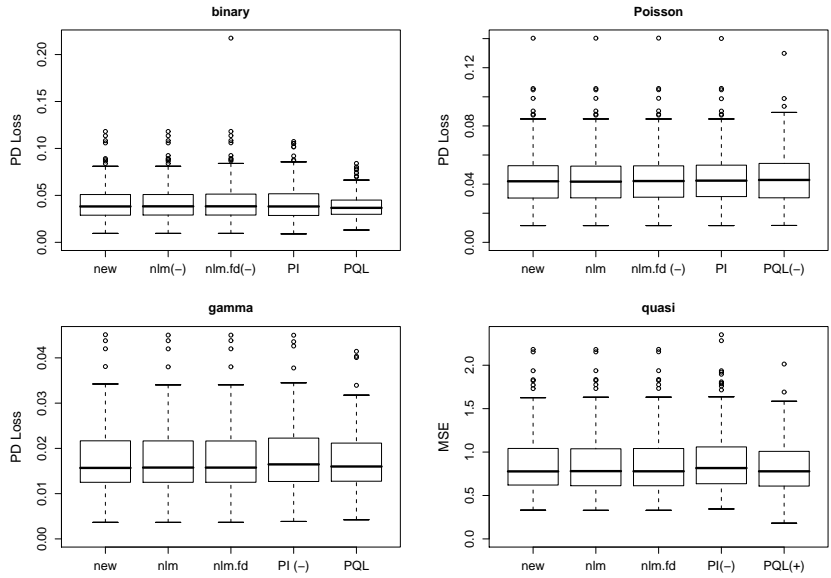}
\vspace*{-.5cm}
\caption{Boxplots of prediction loss measures for the 4 models used in the section \ref{easy.sim} simulations, for each of 5 fitting methods. Labels are as in figure \ref{timing.fig}. The improved speed and reliability is not at the price of prediction performance. \label{how.it.did}}
\end{figure}

Figure \ref{how.it.did} summarizes the predictive performance of the various estimation methods for the 4 types of model. A `(-)' after the label for a method indicates that the method was significantly worse than the new method in a paired comparison using a Wilcoxon signed rank test (using the .05 level); a `(+)' indicates a significant improvement over the new method. The only {\em operationally} significant differences are the worse performance of performance oriented iteration in the gamma and quasi cases, the worse performance of PQL in the Poisson case, and the better performance of PQL in the quasi case, but even these differences are rather modest. Note that the PQL failures were mostly for replicates producing quite high MSE or PD loss results by the other methods (which needs to be born in mind when viewing figure \ref{how.it.did}.) So the new method appears to greatly improve speed and reliability without sacrificing prediction performance.

\subsection{Generalized Additive Mixed Models \label{gamm.sim}}

The same argument that allows a GAM to be estimated as a Generalized Linear Mixed Model using PQL implies that many GLMMs can be estimated by the method developed in this paper. To illustrate this the simulations in the previous section were modified by splitting the data into 40 groups of size 10, and redefining the unscaled linear predictor as $\tilde \eta_i = f_1(x_{1i})+f_2(x_{2i})+f_3(x_{3i} + b_j$ if observation $i$ is from group $j$. The $b_j$ are i.i.d. $N(0,2^2)$ random deviates. The models fitted to each replicate were modified to GAMMs with linear predictors,
$$
g\{E(y_i)\} = f_1(x_{1i})+f_2(x_{2i})+f_3(x_{3i}) + f_4(x_{4i}) + b_j {\rm ~if~}i~{\rm ~from~group~}j.
$$
where the $b_j$ are assumed i.i.d. $N(0,\sigma^2_b)$. The simulations were otherwise un-modified except that only the new method and PQL were compared. For the new method the random effects are treated just like a smooth with the identity matrix as the penalty coefficient matrix, and the associated smoothing parameter controlling $\sigma^2_b$.

\begin{figure}
\eps{-90}{.4}{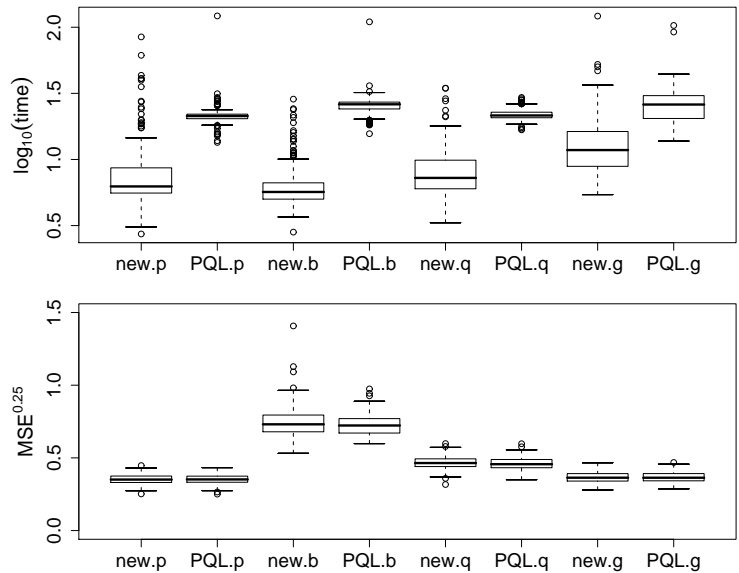}

\vspace*{-.5cm}

\caption{Upper: Boxplots of the distribution of the $\log_{10}$(CPU seconds) used to fit the models to the simulated data in section \ref{gamm.sim}. In the labels {\tt p}, {\tt b}, {\tt q} and {\tt g} refer to the Poisson, binomial, quasi and gamma distributional assumptions. Lower: Boxplots of MSE$^{.25}$ for the various GAMM fits to simulated data discussed in section \ref{gamm.sim}. The new method is faster and more reliable at little or no performance cost. \label{gamm.fig}}
\end{figure}

The mean square error in predicting $\eta_i$ with the $b_j$ set to zero, was used as the measure of model performance, and the number of CPU seconds needed for model estimation was recorded. Out of 200 replicates PQL failed in 22, 12, 16 and 12 replicates for the binary, Poisson, gamma and quasi cases, respectively. The new method did not fail. Timings and MSE performances are shown in figure \ref{gamm.fig}. Wilcoxon tests (paired) fail to detect significant differences between the MSE performance of the methods ($p>.4$), except in the quasi case ($p < 10^{-4}$), where penalized quasi likelihood is significantly better than the new method, perhaps unsurprisingly. Operationally, the MSE differences seem to be small, while the improvements of the new method in terms of speed and reliability are substantial.

\subsection{Severe concurvity \label{concurvity}}

Using R 2.4.0, data were simulated with severe concurvity problems.
\begin{verbatim}
set.seed(23);n <- 400;x <- runif(n);z <- runif(n)
d <- x^3 + rnorm(n)*0.01;f <- (d-.5 + 10*(d-.5)^3)*10
g<-binomial()$linkinv(f);y <- rbinom(g,1,g)
\end{verbatim}
See figure \ref{concurvity.data}. These data are rather extreme, but concurvity problems of this sort are not uncommon in real data examples with many predictors, although they are usually less obvious. The advantage of a simple simulated example is that the root cause of the associated fitting problems is clear, while the `right answer' is known.

\begin{figure}
\eps{-90}{.4}{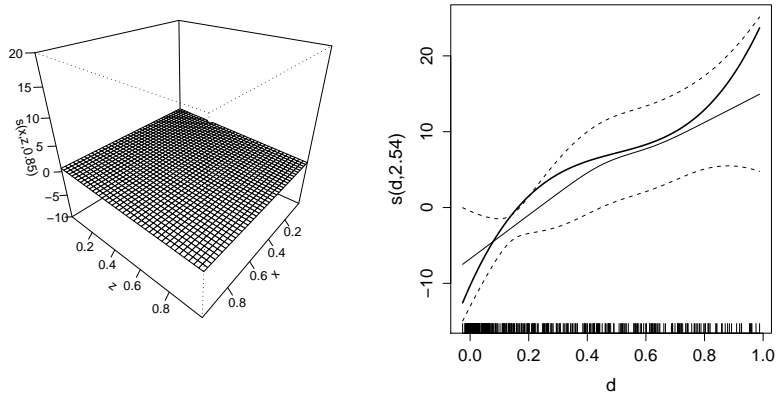}

\vspace*{-.5cm}

\caption{ The estimates of $f_1$ (left) and $f_2$ (right, estimates and 95\% confidence limits) from (\ref{conc.model}), obtained using the new method. $\hat f_1$ has been almost shrunk to zero. The right hand figure also shows the true $f_2$ as a thick black curve (centered in the same way as the smooth). Previous methods fail for this example.
\label{concurvity.fit}}
\end{figure}

The data were modeled using (\ref{conc.model}) as described in the introduction, and existing methods fail to provide satisfactory fits, if they produce a fit at all. The new method converged to an estimated model which substantially suppressed $f_1$ (its effective degrees of freedom were reduced to .8) while doing a reasonable job at estimating $f_2$. The estimated model components are shown in figure \ref{concurvity.fit}. Following Wood (2004) the performance oriented iteration can be made convergent by regularizing the working penalized linear models at each iterate. However the results are very sensitive to the exact degree of regularization performed, with only a narrow window between convergence failure and gross oversmoothing. This is clearly unsatisfactory.

In replicate simulations of this sort the new method is persistently more reliable than the alternatives, although there are of course a substantial number of replicates where all methods perform reasonably (the given replicate is unusual in that {\em all} the alternatives perform badly). No replicates where found where the new method performed less well than the alternatives.

\subsection{A fisheries survey \label{mack.section}}

\begin{figure}
\eps{-90}{.5}{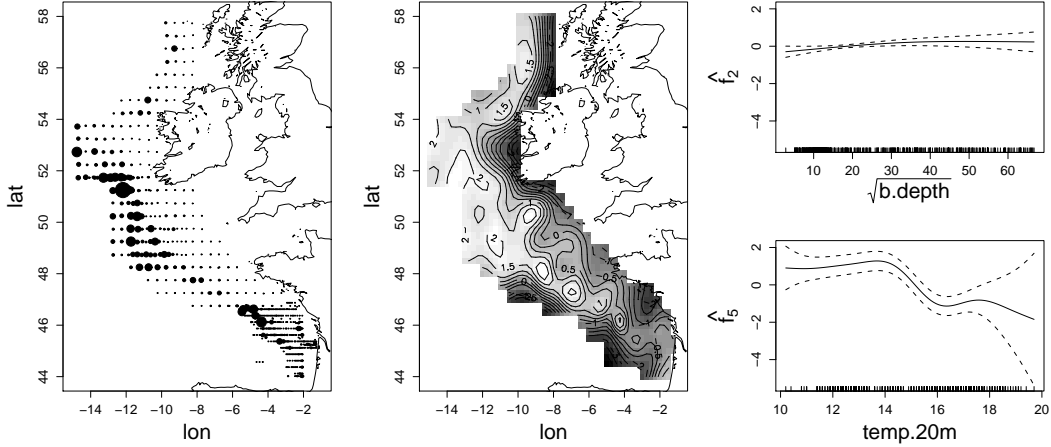}

\vspace*{-.5cm}

\caption{The raw mackerel data (left, symbol area proportional to egg density) and the non-zero estimated terms from the mackerel egg model of section \ref{mack.section}. The central figure shows the spatial smooth over the survey region. The right hand figures show the estimated smooths of square root of sea bed depth and water temperature at 20 metres depth. PQL and performance oriented iteration fail for this example.  \label{mack.fig}}
\end{figure}

This final example concerns modelling of fish egg data from a survey of mackerel eggs conducted in 1992 off the west coast of the British Isles and France. The purpose of the survey is to assess the abundance of fish eggs in order to infer the total mass of spawning fish producing them. The data consist of egg counts from samples collected by hauling a sampling net through the water column from below the maximum depth at which eggs are found, to the surface, and are shown in figure \ref{mack.fig}. Along with egg densities, {\tt egg}, the covariates {\tt long}, {\tt lat}, {\tt b.depth}, {\tt c.dist}, {\tt temp.surf} and {\tt temp.20m} were recorded, these being longitude, latitude, depth of sea bed below surface, distance from the 200m sea bed depth contour (a proxy for distance from the continental shelf edge), surface water temperature and water temperature at 20 m depth, respectively. In addition the area of the sampling net was recorded. There are 634 egg counts spread over the survey area. See Borchers et al. (1997) or Bowman and Azzalini (1997) for further details.

This survey also formed the basis for the presence absence data in figure \ref{mack.bin.data}, used to illustrate  convergence failure in the introduction. Unlike previous methods, the new method successfully fits (\ref{mack.logistic}), identifying a genuine `minimum AIC' model (i.e. the AIC has zero gradient and positive definite Hessian at the optimum). One can make the same point by modelling presence absence over the whole survey area, but given the spatial distribution of presences such a model is not practically defensible.

Turning to the modelling of egg densities (and neglecting any zero inflation problems), a reasonable initial model for the data is
\begin{multline*}
\log\{E({\tt egg}_i)\} = f_1({\tt long}_i,{\tt lat}_i) + f_2(\sqrt{{\tt b.depth}_i}) + f_3({\tt c.dist}_i) + f_4({\tt temp.surf}_i)\\ + f_5({\tt temp.20m}_i) + \log({\tt net.area}_i)
\end{multline*}
along with the `quasi-Poisson' assumption ${\rm var}({\tt egg}_i) \propto E({\tt egg}_i)$ and the assumption that the response variable is independent (conditional on the covariates). $f_1, \ldots, f_5$ can be represented using penalized thin plate regression splines with shrinkage (see Wood, 2006), employing basis dimensions of 100 for $f_1$ and 10 for each of the remaining terms.

Attempts to fit this model using performance oriented iteration fail, without extra regularization: the iteration cycles without ever converging. PQL is no more successful: it diverges until the routine for estimating the working linear mixed model fails. In contrast the new method fits the model without difficulty. The raw fit shows signs of overfitting (informal significance measures for several terms indicate that they have no real effect on the response, despite having fairly high estimated effective degrees of freedom). For this reason the model was re-fitted with $\gamma=1.4$ in the GCV score (see Kim and Gu, 2004). Two model terms were then estimated to have zero effective degrees of freedom (i.e. were penalized out of the model). The remaining terms are shown in figure \ref{mack.fig}.

The difficulties in estimating the model by performance oriented iteration or PQL are again likely to relate to concurvity issues: all the covariates are functions of spatial location, some of them quite smooth functions. In addition the data contain 265 zeroes, and over half the counts are 0 or 1. At these very low counts the assumptions underlying PQL are likely to be somewhat poor, while the linearized problem used in performance oriented iteration is unlikely to capture the full model's dependency on smoothing parameters very precisely.

\section{Conclusions}

Relative to PQL or performance oriented iteration the new method offers two substantial advantages for GAM  (or GAMM) estimation and smoothness selection.
\begin{enumerate}
\item It is more computationally reliable. Since smoothing parameter is based on optimizing a properly defined function, fitting does not suffer from the convergence problems suffered by PQL or performance oriented iteration.
\item The value of the optimized smoothness selection criteria (GCV/AIC) is useful for model comparisons, since it relates to the model being fitted, rather than to some working approximation as is the case for PQL or POI.
\end{enumerate}
In addition the new method is much quicker than PQL, and competitive with performance oriented iteration (in simulations the median cost of the new method is higher while the mean cost is lower). Another less obvious benefit of the new approach is that it integrates easily with step reduction procedures for stabilizing the P-IRLS algorithm if it diverges, as it occasionally does, particularly in the early steps of fitting binary data. Since the P-IRLS is run to convergence with fixed smoothing parameters, it is easy to detect divergence --- this is not the case with performance oriented iteration or PQL, where the smoothing parameters change alongside the parameter estimates at each step of the iteration, so that any possible measure of fit may legitimately increase or decrease from one iteration step to the next. The disadvantage of the new method is the complexity of sections \ref{B.deriv}, \ref{trA.deriv} and associated appendices, with little carrying over from the linear problem. However, this disadvantage is a one off. Once the method has been implemented, it is hard to imagine circumstances in which performance oriented iteration or a finite differencing based method would be preferable.

Relative to finite difference based optimization of GCV/AIC scores, the new method offers much improved computational speed. In difficult modelling situations it also offers enhanced reliability, by elimination of the finite difference approximation error which can lead to false convergence. It is not hard to see why problems might arise in finite differencing. The quantities being differentiated are the converged state of an iterative algorithm, which has to adaptively cope with ill-conditioning problems. Unless very elaborate finite difference schemes are applied there is always a danger that the values that get differenced result from different numbers of steps of the P-IRLS, or have had different levels of truncation applied to cope with ill-conditioning: either case can easily cause the finite difference approximation to fail even to get the sign of the derivative right. The new method eliminates this issue.

An obvious alternative to section \ref{fit.details} would be to use auto-differentiation to automatically accumulate derivatives of the smoothness criteria directly from the computer code evaluating the criteria (see Skaug and Fournier, 2006, for a good statistically based introduction). However, `forward mode' auto-differentiation has an operations count of the same order as finite differencing making it uncompetitive here, while the alternative `reverse mode' requires storage of every intermediate result in the algorithm being differentiated, which is impractical in the current context.

How far does the proposed method go towards the aim, stated in the introduction, of making GAM fitting with smoothness selection as routine as GLM fitting? The aim is the same as that given in Wood (2004), but that paper was restricted to performance oriented iteration, a method for which convergence to any sort of fixed point can not be guaranteed (and may have to be forced by ad hoc regularization). By taking the direct approach the new method is based on optimizing criteria which have well defined optima for any model. This avoids the convergence issue, but replaces it with the problem of how to find the optimum in as efficient and stable a manner as possible, something that is made difficult by the additional non-linearities introduced by the direct approach. The new method succeeds in providing very efficient direct calculation of the derivatives of the smoothness selection criteria, as is evident in the surprising timing results, relative to performance oriented iteration, given in section \ref{easy.sim}. It is unlikely that further substantial improvements are possible in this regard. As highlighted in the introduction, numerical stability is an important and unavoidable issue when working with models as flexible as GAMs, and the methods proposed here directly address the rank deficiency that may cause this. The QR approach to the basic fitting problem is the most stable method known, while the approach taken to rank determination has performance close to the `gold standard' of SVD (see Golub and van Loan, 1996). Again then, there is no obvious alternative that might result in a more stable method. In short, the proposed method achieves the stated aim as closely as is likely to be achievable (which seems to be quite close).

The method described here is implemented in R package {\tt mgcv} (\verb+cran.r-project.org+).

\section*{Acknowledgements}
I am grateful to Stefan Lang for a good deal of help in understanding the issues surrounding GAMs and Bayesian MCMC computation and for help getting started with BayesX, and to the R core for providing the statistical computing environment that made this work a feasible undertaking. I would also like to thank two referees for helpful suggestions on the structure of the paper, and for some other comments which I think have improved it.

\section*{Appendix A: The derivatives of $\bf z$}

The $\bf z$ derivative update referred to in section \ref{section.pirls} is given here. Note that $\mu_i$, $\eta_i$, $z_i$ and $w_i$ are always taken as being evaluated at the converged $\hat \bp$.

\noindent {\bf Initialization:}
${\bf z}$, $\bf w$, $\bm \mu$ and $\bm \eta$ are fixed at their converged values from the P-IRLS, but all their derivatives w.r.t. $\bm \rho$ are initially set to zero. The initial derivatives of $\hat {\bm \beta}$ w.r.t. $\bm \rho$ are as in section \ref{section.pirls}.
At the converged estimate of $\mu_i$, evaluate the constants: $c_{1i}=(y_i - \mu_i) g^{\prime\prime}(\mu_i)/g^\prime(\mu_i)$, $c_{2i} = [(y_i-\mu_i) \{ g^{\prime\prime\prime}(\mu_i)/g^{\prime}(\mu_i)^2
- g^{\prime\prime}(\mu_i)^2/g^{\prime}(\mu_i)^3
\} - g^{\prime\prime}(\mu_i)/g^{\prime}(\mu_i)^2 ]$,
$c_{3i} = w_i^3\{V^\prime(\mu_i)g^\prime(\mu_i) + 2 V(\mu_i) g^{\prime\prime}(\mu_i) \}/(2 \omega_i)$ and
$c_{4i} = w_i^3\{
V^{\prime\prime}(\mu_i)g^\prime(\mu_i)  + 2 g^{\prime \prime\prime}(\mu_i)V(\mu_i) + 3 g^{\prime\prime}(\mu_i) V^{\prime}(\mu_i)
\} / \{2 \omega_i{g^\prime(\mu_i)}\}$.

\noindent {\bf Update:}
The following steps update the $\bf z$ derivatives given the $\hat \bp$ derivatives (for all $k,m$, such that $k \ge m$).
\begin{enumerate}

\item Evaluate
$$
\pdif{\bm \eta}{\rho_k} = \X \pdif{\hat \bp}{\rho_k} ~~~{\rm and}~~~ \pddif{\bm \eta} {\rho_k}{\rho_m} = \X \pddif{\hat \bp}{\rho_k}{\rho_m}.
$$
\item Update the derivatives of ${\bf z}$:
$$
\pdif{z_i}{\rho_k} =  c_{1i}\pdif{\eta_i}{\rho_k} {\rm ~~and~~}
\pddif{z_i}{\rho_k}{\rho_m}
= c_{1i} \pddif{\eta_i^2}{\rho_k}{\rho_m}
+ c_{2i} \pdif{\eta_i}{\rho_m} \pdif{\eta_i}{\rho_k}.
$$
\item Update the derivatives of $w_i = \omega_i^{1/2}V(\mu_i)^{-1/2}/g^{\prime}(\mu_i)$:
$$
\pdif{w_i}{\rho_k} = -
c_{3i}
\pdif{\eta_i}{\rho_k}
{\rm ~~and~~}
\pddif{w_i}{\rho_k}{\rho_m} = \frac{3}{ w_i} \pdif{w_i}{\rho_k} \pdif{w_i}{\rho_m}
-
c_{3i}
\pddif{\eta_i}{\rho_k}{\rho_m}
-
c_{4i}
\pdif{\eta_i}{\rho_m}\pdif{\eta_i}{\rho_k}.
$$
\item The derivatives of ${\bf z}^{\prime}$ are evaluated:
$$
\pdif{z_i^\prime}{\rho_k} = \pdif{w_i}{\rho_k}z_i + w_i \pdif{z_i}{\rho_k}
{\rm ~~and~~}
\pddif{z_i^\prime}{\rho_k}{\rho_m} = \pddif{w_i}{\rho_k}{\rho_m} z_i  + \pdif{w_i}{\rho_k}\pdif{z_i}{\rho_m} + \pdif{w_i}{\rho_m}\pdif{z_i}{\rho_k} + w_i\pddif{z_i}{\rho_k}{\rho_m}.
$$

\end{enumerate}

\section*{Appendix B: Deviance and Pearson statistic derivatives}

The derivatives of the deviance can be obtained as follows.
$$
\pdif{D}{\rho_k} = \sum_j\pdif{D}{\hat \beta_j}\pdif{\hat \beta_j}{\rho_k} {\rm ~~and~~}
\pddif{D}{\rho_k}{\rho_m} = \sum_j \left (\sum_l \pddif{D}{\hat \beta_j}{\hat \beta_l} \pdif{\hat \beta_l}{\rho_m} \pdif{\hat \beta_j}{\rho_k}\right ) + \pdif{D}{\hat \beta_j} \pddif{\hat \beta_j}{\rho_k}{\rho_m}.
$$

The required derivatives of the deviance w.r.t. $\hat \bp$ are
$$
\pdif{D}{\hat \beta_j} = -2 \sum_i \omega_i \frac{y_i - \mu_i}{V(\mu_i)g^\prime(\mu_i)} X_{ij} {\rm ~~and}
$$
$$
\pddif{D}{\hat \beta_j}{\hat \beta_l} = 2 \sum_i \omega_i \left [ \frac{1}{V(\mu_i)g^\prime(\mu_i)} \pdif{\mu_i}{\hat \beta_l} + \frac{y_i - \mu_i}{[V(\mu_i)g^\prime(\mu_i)]^2} \left \{ V^\prime(\mu_i) g^\prime(\mu_i) + V(\mu_i) g^{\prime\prime}(\mu_i) \right \} \pdif{\mu_i}{\hat \beta_l}
 \right ] X_{ij}.
$$

So, defining ${\bf c}$ as the vector with elements
$
c_i = -2 \omega_i (y_i - \mu_i)/\{V(\mu_i)g^\prime(\mu_i) \},
$
the vector of first derivatives of $D$ w.r.t. the $\hat \beta_j$ is $\X\ts {\bf c}$. Now noting that $\ilpdif{\mu_i}{\hat \beta_l} = X_{il}/g^\prime(\mu_i)$ and defining
$$
e_i = 2 \omega_i \left [\frac{1}{V(\mu_i)g^\prime(\mu_i)^2} + \frac{y_i - \mu_i}{V(\mu_i)^2g^\prime(\mu_i)^3} \left \{V^\prime(\mu_i)g^\prime(\mu_i) + V(\mu_i) g^{\prime\prime}(\mu_i) \right \}   \right ],
$$
the second derivative matrix (Hessian) of $D$ is $\X\ts{\rm diag}(e_i)\X$.

The derivatives of the Pearson statistic, $P$, are easily obtained by noting that
$$
P = \sum_{i=1}^n \omega_i \frac{(y_i - \hat \mu_i)^2}{V(\hat \mu_i)} = \sum_{i=1}^2 w_i^2(z_i - \hat \eta_i)^2.
$$
The expression in terms of the iterative weights, pseudodata and linear predictor makes evaluation of the derivatives of $P$ particularly straightforward, since the derivatives of all $w_i$, $z_i$ and $\hat \eta_i$ are available directly from the derivative iteration.

\section*{Appendix C: Efficient evaluation of the derivatives of $\tr{\bf A}$}

In the following, wherever $\sqrt{{\bf S}_m}$ is written it denotes the $q \times {\rm rank}({\bf S}_m)$ matrix such that $\sqrt{{\bf S}_m}\sqrt{{\bf S}_m}\ts={\bf S}_m$ (pivoted Choleski decomposition can be used to find these, see Golub and van Loan, 1996 and Dongarra et al. 1978). The following list gives the key steps for evaluating each of the different types of term making up the second derivatives of $\tr{\bf A}$ as given on the RHS of equation (\ref{trA.2deriv}).
\begin{enumerate}
\item For $\tr{\Tkm \A}$ etc. first form and store $\diag{\A} = \diag{\K\K\ts}$ and the term follows.

\item $\tr{\Tk\A\Tm\A} = \tr{[\K\ts\Tk\K][\K\ts\Tm\K]} = \tr{\Tm\A\Tk\A}$ (the second equality follows from transposing the matrix expression in the middle trace). This requires storage of $\K\ts\Tk\K$, in advance.

\item
Terms like $\tr{\A\Tkm\A}$ follow from $\tr{\A\Tkm\A}=\tr{\Tkm\A\A}$. So, $\diag{\A\A} =
\diag{[\K \K \ts \K][\K \ts]}$ is evaluated once up front (having first formed $\K\ts \K $ and then $\K \K \ts \K  $) and the result is then readily computed.

\item $\K\ts\Tk\K\K\ts\K$ is stored up front so that use can be made of\\
$\tr{\A\Tk\A\Tm\A}=\tr{[\K\ts\Tk\K][\K\ts\Tm\K\K\ts\K]}$.

\item
$\tr{\A\Tk\B\ts{\bf S}_m\B}= \tr{\Tk[\K\p\ts\sqrt{{\bf S}_m}][\sqrt{{\bf S}_m}\ts\p\K\ts\K\K\ts]}$, so evaluate \\
$\diag{[\K\p\ts\sqrt{{\bf S}_m}][\sqrt{{\bf S}_m}\ts\p\K\ts\K\K\ts}$ and the result is easily obtained.
This required up front storage of $\K\K\ts\K\p\ts\sqrt{{\bf S}_m}$ and $\K\p\ts\sqrt{{\bf S}_m}$.

\item
Evaluate $\diag{\B\ts{\bf S}_m \B} = \diag{[\K\p\ts\sqrt{{\bf S}_m}][\sqrt{{\bf S}_m}\ts\p\K\ts]}$ and terms like \\ $\tr{\Tk\B\ts{\bf S}_m\B}$ follow easily.

\item Finally, if $\p\ts{\bf S}_m \p$ and $\p\ts{\bf S}_m \p \K \ts \K$ are stored up front, then\\
$\tr{\B\ts {\bf S}_m \Gi {\bf S}_k\B} = \tr{[\p\ts{\bf S}_m \p][\p\ts{\bf S}_k\p\K\ts\K]}$ is easily obtained.
\end{enumerate}
Notice that, if $M$ is the number of smoothing parameters, then by far the most expensive calculation here is the evaluation of the $M$ terms $\K\ts\Tk\K$ in step 2. This has a total cost of $nq^2M/2$ floating point operations, which is still a considerable saving over finite differencing to get second derivatives. Note also that all terms in $\tr{\bf A} $ and its first derivatives are covered in the above list, and have a total leading order computational cost of $O(nq^2)$, the same as model estimation: this is an $M+1$ fold saving over finite differencing.

\subsection*{References}

\begin{trivlist}
\item Anderson, E., Z. Bai, C. Bischof, S. Blackford, J. Demmel, J. Donngarra, J. Du Croz, A. Greenbaum,
S. Hammerling, A. McKenney \& D. Sorenson (1999) {\em LAPACK Users' Guide} (3rd ed.) SIAM, Philadelphia.

\item Akaike, H. (1973) Information theory and an extension of the maximum likelihood principle. In B. Petran \& F. Csaaki (Eds.) {\em International Symposium on Information Theory}, Akadeemiai Kiadi, Budapest, Hungary, pp. 267-281.

\item Borchers, D.L., S.T. Buckland, I.G. Priede \& S. Ahmadi (1997) Improving the precision of the daily egg production method using generalized additive models. {\em Canadian Journal of fisheries and aquatic science} 54, 2727-2742.

\item Bowman, A.W. \& A. Azzalini (1997) {\em Applied Smoothing Techniques for Data Analysis} Oxford University Press.

\item Breslow, N.E. \& D.G. Clayton (1993) Approximate inference in generalized linear mixed models. {\em Journal of the American Statistical Association} 88, 9-25.

\item Brezger, A. \& S. Lang (2006) Generalized structured additive regression based on Bayesian P-splines. {\em Computational Statistics and Data Analysis} 50, 967-991.

\item Brezger, A., T Kneib \& S. Lang (April 2007) {\em BayesX} 1.5.0 \verb+http://www.stat.uni-muenchen.de/~bayesx+

\item Cline, A.K, C.B. Moler, G.W. Stewart \& J.H. Wilkinson (1979) An Estimate for the Condition Number of a Matrix. {\em SIAM Journal of Numerical Analysis} 13, 293-309.

\item Craven P. \& G. Wahba (1979) Smoothing noisy data with spline functions: Estimating the correct degree of smoothing by the method of generalized cross validation {\em Numerische Mathematik} 31, 377-403.

\item Dennis, J. E. \& R.B. Schnabel (1983) {\em Numerical Methods for
     Unconstrained Optimization and Nonlinear Equations.}
     Prentice-Hall, Englewood Cliffs, NJ.

\item Dongarra, J. J., J.R Bunch, C.B. Moler \& G.W. Stewart (1978) {\em LINPACK Users Guide} SIAM, Philadelphia.

\item Eilers, P.H.C. \& B.D. Marx (2002) Generalized linear additive smooth structures. {\em Journal of computational and graphical statistics} 11(4), 758-783.

\item Fahrmeir, L. T. Kneib \& S. Lang (2004) Penalized structured additive regression for space time dataL: A Bayesian perspective. {\em Statistica Sinica} 14, 731-761.

\item Fahrmeir, L. \& S. Lang (2001) Bayesian inference for generalized additive mixed models based on Markov random field priors. {\em Applied Statistics} 50, 201-220

\item Figueiras, A., J. Roca-Pardi\~nas \& C. A. Cadarso-Su\'arez (2005) A bootstrap method to avoid the effect of concurvity in generalized additive models in time series studies of air pollution. {\em Journal of Epidemiology and Community Health} 59, 881-884.

\item Gill, P.E., W. Murray \& M.H. Wright (1981) {\em Practical Optimization} Academic Press, London.

\item Golub, G.H. \& C.F. van Loan (1996) {\em Matrix Computations} (3rd edition). Johns Hopkins University Press, Baltimore.

\item Green, P.J. \& B.W. Silverman (1994) {\em Nonparametric Regression and Generalized Linear Models} Chapman \& Hall, London.

\item Gu, C. \& G. Wahba (1991) Minimizing GCV/GML scores with multiple smoothing parameters via the Newton method. {\em SIAM Journal on Scientific and Statistical Computing} 12, 383-398.

\item Gu, C. (1992) Cross validating non-Gaussian data. {\em Journal of Computational and Graphical Statistics} 1, 169-179.

\item Gu, C. (2002) Smoothing Spline ANOVA Models. Springer, New York.

\item Gu, C. (2004) {\tt gss}: General Smoothing Splines. R package version 0.9-3.

\item Gu, C. \& D. Xiang (2001) Cross-validating non-Gaussian data: Generalized approximate cross-validation revisited. {\em Journal of Computational and Graphical Statistics} 10, 581-591.

\item Hastie, T. \& R. Tibshirani (1986) Generalized additive models (with discussion). {\em Statistical Science} 1, 297-318.

\item Hastie, T. \& R. Tibshirani (1990) {\em Generalized additive models} Chapman \& Hall, London.

\item Hastie, T. \& Tibshirani (1993) Varying-coefficient models. {\em Journal of the Royal Statistical Society, Series B} 55, 757-796.

\item Kim, Y.J. \& Gu, C. (2004) Smoothing spline Gaussian regression: more scalable computation via efficient approximation. {\em Journal of the Royal Statistical Society, Series B} 66, 337-356.

\item Lang, S \& A. Brezger (2004) Bayesian P-splines. {\em Journal of Computational and Graphical Statistics} 13, 183-212

\item Lin, X. \& D. Zhang (1999) Inference in generalized additive mixed models using smoothing splines. {\em Journal of the Royal Statistical Society, Series B} 61, 381-400.

\item Mallows, C.L. (1973) Some comments on $C_p$ {\em Technometrics} 15, 661-675.

\item Marx B. D. \& P.H. Eilers (1998) Direct generalized additive modeling with penalized likelihood. {\em Computational Statistics and Data Analysis} 28, 193-209.

\item McCullagh P. \& J. A. Nelder (1989) Generalized linear models (2nd ed.) Chapman \& Hall, London.

\item Nelder, J.A. \& R.W.M. Wedderburn (1972) Generalized linear models. {\em Journal of the Royal Statistical Society, Series A} 135, 370-384

\item O'Sullivan, F. (1986) A statistical perspective on ill-posed inverse problems.  {\em Statistical Science} 1, 502-518

\item O'Sullivan, F.B., B. Yandall \& W. Raynor (1986) Automatic smoothing of regression functions in generalized linear models. {\em Journal of the American Statistical Association} 81, 96-103.

\item Pinheiro, J.C. \& D.M. Bates (2000) {\em Mixed-Effects Models in S and S-PLUS} Springer, New York.

\item R Core Development Team (2006) R 2.4.0: A Language and Environment for Statistical Computing. R Foundation for Statistical Computing, Vienna.

\item Ramsay, T., R. Burnett \& D. Krewski (2003) Exploring bias in a generalized additive model for spatial air pollution data. {\em Environmental Health Perspectives} 111, 1283-1288

\item Ruppert, D., M.P. Wand \& R.J. Carroll (2003) {\em Semiparametric Regression} Cambridge University Press.

\item Skaug, H.J. \& D. Fournier (2006) Automatic approximation of the Marginal Likelihood in non-Gaussian Hierarchical Models. {\em Computational Statistics and Data Analysis} 51, 699-709

\item Stone, M. (1977) An asymptotic equivalence of choice of model by cross-validation and Akaike's criterion. {\em Journal of the Royal Statistical Society, Series B} 39, 44-47.

\item Wahba, G (1990) {\em Spline models for observational data} SIAM, Philadelphia.

\item Watkins, D.S. (1991) {\em Fundamentals of Matrix Computations}  Wiley, New York.

\item Wood, S.N. (2000) Modelling and smoothing parameter estimation with multiple quadratic penalties.
{\em Journal of the Royal Statstical Society, Series B} 62, 413-428.

\item Wood, S.N. (2003) Thin plate regression splines. {\em Journal of the Royal Statistical Society, Series B} 65, 95-114.

\item Wood, S.N. (2004) Stable and efficient multiple smoothing parameter estimation for generalized additive models. {\em Journal of the American Statistical Association} 99, 673-686.

\item Wood, S.N. (2006) {\em Generalized Additive Models: An Introduction with R} CRC/Chapman \& Hall, Boca Raton, Florida.

\item Xiang, D. \& G. Wahba (1996) A generalized approximate cross validation for smoothing splines with non-Gaussian data. {\em Statistica Sinica} 6, 675-692.
\end{trivlist}

 \end{document}